\documentclass[twocolumn]{aastex631}
\newcommand{\vlos}{\ensuremath{v_\mathrm{LOS}}}
\newcommand{\vabs}{\ensuremath{v_\mathrm{abs}}}

\newcommand{\unitpw}[2]{\ensuremath{\mathrm{#1}^{#2}}}
\newcommand{\per}[1]{\unitpw{#1}{-1}}

\newcommand{\percb}[1]{\unitpw{#1}{-3}}
\newcommand{\kmps}{\ensuremath{\mathrm{km}\,\per{s}}}

\defcitealias{W16}{W16}
\defcitealias{P1}{I}
\defcitealias{P2}{II}
\defcitealias{P3}{III}
\defcitealias{P4}{IV}
\defcitealias{P5}{V}
\defcitealias{M22}{M22}
\defcitealias{I23}{I23}

\received{2024 January 19}
\revised{2024 March 11}
\accepted{2024 Marcn 14}

\submitjournal{ApJ}

\shorttitle{Molecular inflow and atomic outflow in Circinus}
\shortauthors{Baba et al.}

\graphicspath{./}

\begin{document}

%% Title
\title{Circumnuclear Multi-phase Gas in the Circinus Galaxy. VI. Detectability of Molecular Inflow and Atomic Outflow}
%% You can use \\ to force a line break and include a footnote in the title.

%% Authors
%% The \author command takes the 16 digit ORCID as an optional argument
%% \author[xxxx-xxxx-xxxx-xxxx]{Author Name}
%% The \affiliation command automatically indexes affiliations in the header.
%% Multiple calls of \affiliation documents more than one affiliation.
%% The \altaffiliation command indicates some secondary information such as fellowships.
%% This command must come BEFORE the \affiliation call, right after the \author command.
%% The \email command provides email addresses with one call.
%% The \correspondingauthor command identifies the corresponding author.
%% It is the author's responsibility to make sure this name is also in the author list.

\correspondingauthor{Shunsuke Baba}
\email{shunsuke.baba@astrophysics.jp}

\author[0000-0002-9850-6290]{Shunsuke Baba}
\affiliation{Graduate School of Science and Engineering, Kagoshima University, 1-21-35 Korimoto, Kagoshima, Kagoshima 890-0065, Japan}
\affiliation{Institute of Space and Astronautical Science (ISAS),
Japan Aerospace Exploration Agency (JAXA),
3-1-1 Yoshinodai, Chuo-ku, Sagamihara, Kanagawa 252-5210, Japan}

\author[0000-0002-8779-8486]{Keiichi Wada}
\affiliation{Graduate School of Science and Engineering, Kagoshima University, 1-21-35 Korimoto, Kagoshima, Kagoshima 890-0065, Japan}
\affiliation{Research Center for Space and Cosmic Evolution, Ehime University, 2-5 Bunkyo-cho, Matsuyama, Ehime 790-8577, Japan}
\affiliation{Faculty of Science, Hokkaido University, N10 W8, Kita-ku, Sapporo, Hokkaido 060-0810, Japan}

\author[0000-0001-9452-0813]{Takuma Izumi}
\affiliation{National Astronomical Observatory of Japan (NAOJ), 2-21-1 Osawa, Mitaka, Tokyo 181-8588, Japan}
\affiliation{Department of Astronomical Science, The Graduate University for Advanced Studies (SOKENDAI), 2-21-1 Osawa, Mitaka, Tokyo 181-8588, Japan}

\author[0000-0003-0548-1766]{Yuki Kudoh}
\affiliation{Astronomical Institute, Graduate School of Science, Tohoku University, 6-3 Aramaki, Aoba-ku, Sendai, Miyagi 980-8578, Japan}

\author[0000-0002-5012-6707]{Kosei Matsumoto}
\affiliation{Sterrenkundig Observatorium, Universiteit Gent,
Krijgslaan 281 S9, B-9000 Gent, Belgium}
\affiliation{Department of Physics, Graduate School of Science,
The University of Tokyo,
7-3-1 Hongo, Bunkyo-ku, Tokyo 113-0033, Japan}
\affiliation{Institute of Space and Astronautical Science (ISAS),
Japan Aerospace Exploration Agency (JAXA),
3-1-1 Yoshinodai, Chuo-ku, Sagamihara, Kanagawa 252-5210, Japan}

%% Abstract
%% A single paragraph of not more than 250 words, except for Research Notes (150 words).
\begin{abstract}
Recent submillimeter observations have revealed signs of pc-scale molecular inflow and atomic outflow in the nearest Seyfert 2 galaxy, the Circinus galaxy.
To verify the gas kinematics suggested by these observations, we performed molecular and atomic line transfer calculations based on a physics-based 3D radiation-hydrodynamic model, which has been compared with multi-wavelength observations in this paper series.
The major axis position--velocity diagram (PVD) of CO(3--2) reproduces the observed faint emission at the systemic velocity, and our calculations confirm that this component originates from failed winds falling back to the disk plane.
The minor-axis PVD of [\ion{C}{1}]($^3P_1$--$^3P_0$), when created using only the gas with positive radial velocities, presents a sign of blue- and redshifted offset peaks similar to those in the observation, suggesting that the observed peaks indeed originate from the outflow, but that the model may lack outflows as strong as those in the Circinus galaxy.
Similar to the observed HCN(3--2), the similar dense gas tracer HCO$^+$(3--2) can exhibit nuclear spectra with inverse P-Cygni profiles with $\sim$0.5\,pc beams, but the line shape is azimuthally dependent.
The corresponding continuum absorbers are inflowing clumps at 5--10\,pc from the center.
To detect significant absorption with a high probability, the inclination must be fairly edge-on ($\gtrsim$85$^\circ$), and the beam size must be small ($\lesssim$1\,pc).
These results suggest that HCN or HCO$^+$ and [\ion{C}{1}] lines are effective for observing pc-scale inflows and outflows, respectively.
\end{abstract}

%% Keywords
\keywords{Active galactic nuclei(16) --- Seyfert galaxies(1447) --- Interstellar medium(847) --- Radiative transfer simulations(1967) --- Millimeter astronomy(1061) --- Submillimeter astronomy(1647)}

%% Body

\section{Introduction} \label{sec:intro}

The gas dynamics in the vicinity of active galactic nuclei (AGNs) are crucial for understanding the coevolution of supermassive black holes (SMBHs) and their host galaxies.
Outflows from AGNs play an important role in the negative feedback on massive galaxies \citep{Kormendy&Ho13} and are observed in the molecular, atomic, and ionized phases \citep[e.g.,][]{Fluetsch+19,Morganti+05,Muller-Sanchez+11}.
Inflows toward AGNs, on the other hand, are responsible for the feeding or mass accretion of black holes and are observed mainly in the cold molecular phase \citep[e.g.,][]{Veilleux+13,Tremblay+16,Rose+23}.
How these gas flows in the different phases occur and how they can be observed at the pc scale remain to be elucidated.
In particular, observational understandings of the driving processes of inflows have been accumulated down to the 100\,pc scale 
\citep{Storchi-Bergmann&Schnorr-Muller19}, but there have been virtually no observations that spatially resolve the nuclear few-pc region, where SMBHs dominate the gas dynamics.

Recently, \citet[][hereafter \citetalias{I23}]{I23} observed one of the nearest Seyfert 2 galaxies, the Circinus galaxy ($D=4.2$\,Mpc), with a pc-scale resolution using the Atacama Large Millimeter/submillimeter Array (ALMA).
They observed signs of molecular inflow and atomic outflow.
Firstly, the medium-density molecular gas line CO($J$=3--2) (345.796\,GHz, $n_\mathrm{crit}\sim1\times10^4\,\percb{cm}$) showed a faint emission close to the systemic velocity in its major axis position--velocity diagram (PVD).
The authors proposed that this slow component was due to a molecular inflow that was not supported by rotation in the disk plane.
Secondly, the minor axis PVD of the diffuse atomic line [\ion{C}{1}] $^3P_1$--$^3P_0$ (hereafter [\ion{C}{1}](1--0), 492.161\,GHz, $n_\mathrm{crit}\sim4\times10^2\,\percb{cm}$) revealed two peaks offset from the nucleus and shifted from the systemic velocity, which were interpreted as originating from atomic outflows.
Thirdly, the dense-gas tracer HCN($J$=3--2) (265.886\,GHz, $n_\mathrm{crit}\sim6\times10^6\,\percb{cm}$) was detected in the continuum absorption toward the nucleus, and the spectrum exhibited a profile consisting of redshifted absorption and blueshifted emission (i.e., an inverse P-Cygni profile).
This feature was interpreted as direct evidence of an inflowing gas in \citetalias{I23}.

As interferometric facilities are developed for higher resolutions and sensitivities, more pc-scale observations will be made for AGNs, not limited to the one in the Circinus galaxy.
In this context, the results and interpretations of \citetalias{I23} as the first example of such an observation should be verified by comparison with a theoretical model of AGN.
Discussions on the detectability of inflows and outflows based on these comparisons are useful for evaluating future observational strategies.
So far, no comparisons have been made, even qualitative ones.

To describe the pc-scale gas dynamics and chemistry around an AGN, we have developed a 3D numerical model named ``radiation-driven fountain'' and demonstrated its validity in this series of papers.
The basic mechanism of the model is not phenomenological but a combination of various fundamental physical processes, and was first proposed by \citet{Wada12}.
Under the influence of AGN radiation, dusty gas transported from the equatorial plane to the nucleus is blown away by the radiation pressure and gas heating, but the velocity of the majority of it does not exceed the escape velocity.
Instead, it turns to the disk plane (failed winds), which results in a quasi-stationary circulation that forms a geometrically thick disk.
This mechanism explains the geometrically and optically thick structure, the torus, which has been postulated to explain the Type-1 and Type-2 dichotomy of AGNs \citep{Antonucci&Miller85}.
\citet[][hereafter \citetalias{W16}]{W16} extended this model to include the chemistry under the effect of the X-ray (X-ray dominated region) and the energy feedback from supernovae occurring in the disk plane.

In previous studies in this series, a model snapshot from \citetalias{W16} has been tested by post-processed radiative transfer calculations, and the observed spectral energy distribution (SED) of the Circinus galaxy was reproduced for a viewing angle of $75^\circ$ or greater.
\citet{P1} and \citet{P2} (Papers \citetalias{P1} and \citetalias{P2}) performed non-local thermodynamic equilibrium (non-LTE) line transfer calculations of submillimeter molecular (CO) and atomic ([CI]) lines.
\citet[][Paper \citetalias{P3}]{P3} focused on the ionized outflow of the radiation-driven fountain and found that the biconical narrow line region and line ratios are consistent with the observations of Seyfert galaxies.
\citet[][Paper \citetalias{P4}]{P4} calculated the CO lines with the dust emission (details provided below).
\citet[][Paper \citetalias{P5}]{P5} computed the X-ray polarization and compared it with a recent observation of Circinus with the Imaging X-ray Polarimetry Explorer (IXPE).
In addition, \citet[][hereafter \citetalias{M22}]{M22} performed the radiative transfer of near-infrared CO rovibrational lines and investigated the detectability of molecular inflows and outflows in the torus, and \citet{Ogawa+22} predicted the X-ray spectrum and found that it is in agreement with that of Circinus.

However, despite these previous results, no radiative transfer has been made that can be compared with the observations of pc-scale gas flows such as \citetalias{I23}.
In Paper \citetalias{P2}, the [\ion{C}{1}](1--0) line was calculated, but the thermal continuum emission from dust heated by the AGN was not considered, and only the global properties over the entire model (32\,pc size) for the gas kinematics were discussed.
In Paper \citetalias{P4}, the CO rotational lines were calculated while including the dust emission, and the possibility of observing them as absorption lines at the nuclear position was explored;  however, no other lines were investigated and neither was their relationship with the kinematics in the circumnuclear region thoroughly discussed.
Observational studies often discuss kinematics using line PVDs and velocity profiles, as in \citetalias{I23}, but these provide less information than the actual three-dimensional gas distribution and velocity field, and their interpretation is not always conclusive.
To understand the true structure from observations, theoretical predictions of how gas kinematics would be observed in PVDs and velocity profiles should be presented based on the radiative transfer calculations of a 3D hydrodynamic simulation.

In this study, we perform the radiative transfer of CO, [\ion{C}{1}], and HCO$^+$ (a dense gas tracer similar to HCN) lines with the dust continuum based on the radiation-driven fountain model and compare the results with the recently discovered signs of pc-scale molecular inflow and atomic outflow in the Circinus galaxy to verify the interpretation of the observations and discuss the detectability of the inflow and outflow.
We use HCO$^+$ instead of HCN because the model chemistry does not currently include nitrogen (Section \ref{sec:model}).
HCO$^+$ is an appropriate surrogate because it has energy levels close to those of HCN and critical densities of the same order of magnitude as HCN \citep{Shirley15}.\footnote{Based on radiative transfer calculations for a 3D hydrodynamic simulation of an AGN torus, \citet{Yamada+07} found that the HCN-to-HCO$^+$ line ratio at $J=1$--0 is $\lesssim$1 over an abundance range of two orders of magnitude when the abundances of these molecules are equal. On the other hand, the HCN-to-HCO$^+$ line ratios in AGNs are observed to be in the range of $\sim$0.5--3 at $J=2$--1, 3--2, and 4--3, which is attributed to efficient collisional excitation and enhanced HCN abundance in AGNs \citep{Imanishi&Baba+23a,Imanishi&Baba+23b}. In any case, the HCN-to-HCO$^+$ intensity ratio is of the order of unity.}
We discuss gas kinematics using PVDs and velocity profiles.
By tracing the process of radiative transfer calculations, we identify where the PVD patterns and spectral features originate and provide insights that cannot be obtained from observational studies.
% Note that \citetalias{I23} also observed the H36$\alpha$ recombination line, revealed its horn-like morphology, and suggested that it traces the outline of a conical ionized outflow, but this line is not discussed in this paper.
The goal of this paper is to make qualitative comparisons using the existing model snapshot.
Tuning the model to fit the Circinus galaxy better is left for future work because the influence of the model parameters (e.g., SMBH mass and AGN luminosity; Section \ref{sec:model}) on the observables is complicated and not quantitatively predictable, and simulating the model with many different setups is computationally expensive.

The remainder of this paper is organized as follows.
The fountain model and our radiative transfer method are detailed in Section \ref{sec:method}.
The results are presented in Section \ref{sec:results} and discussed in Section \ref{sec:discussion}.
Finally, the conclusions are summarized in Section \ref{sec:conclusions}.

\section{Method} \label{sec:method}

\subsection{Input Radiation-Driven Fountain Model} \label{sec:model}

The input model snapshot used in this study is identical to that used in Papers \citetalias{P1}--\citetalias{P4} and \citetalias{M22}.
It is described briefly in this section.

The radiation-driven fountain model is a 3D grid-based hydrodynamic simulation of the $r\leqslant 16$\,pc region around an AGN.
A rotating gas disk in the gravitational potential evolves under the influence of radiation.
The simulation box comprises a uniform Cartesian grid of $256^3$ cells.
The central AGN is assumed to have a mass of $M_\mathrm{BH}=2\times10^6\,M_\odot$,\footnote{Based on the measurement by \citet{Greenhill+03} from the H$_2$O maser in the Circinus galaxy. \citetalias{I23} obtained a consistent mass by modeling the ALMA observations.} and the total gas mass is set as $M_\mathrm{gas}=2\times10^6\,M_\odot$.\footnote{This mass was determined in \citetalias{W16} to reproduce the observed depth of the 9.7\,\micron\ silicate dust absorption in the Circinus galaxy.}
The dust is included at a uniform dust-to-gas mass ratio of 0.01.
To represent the galactic potential, a time-independent external potential $\Phi_\mathrm{ext}(r)=-(27/4)^{1/2} [v_1^2 a_1 /(r^2+a_1^2)^{1/2} + v_2^2 a_2 / (r^2+a_2^2)^{1/2}]$ is included, where $a_1=100$\,pc, $a_2=2.5$\,kpc, and $v_1=v_2=147\,\kmps$.
The self-gravity of the gas is ignored.

Radiative feedback processes are implemented using ray tracing.
The AGN bolometric luminosity is assumed to be constant at $L_\mathrm{bol}=5\times10^{43}\,$erg\,\per{s} with a fixed Eddington ratio of 0.2.\footnote{Based on the estimate by \citet{Tristram+07} from mid-infrared interferometric observations of the Circinus galaxy.}
The ultraviolet (UV) and X-ray fluxes are calculated from $L_\mathrm{bol}$ \citep{Marconi+04}, and both play an important role in the radiation pressure on the dust and heating of the gas,\footnote{In addition to the X-ray heating, the gas is also heated owing to feedback from the circumnuclear disk via photoelectric heating due to far-UV radiation from stars and energy injection from supernova explosions. These effects are implemented as described in \citetalias{W16}.} respectively.
Both these effects drive the outflows.
The UV flux is assumed to be anisotropic because of the limb darkening of the accretion disk, with a dependence of $F(\theta)\propto\cos\theta(1+2\cos\theta)$, where $\theta$ is the angle from the rotation axis \citep{Netzer+87}.
The dust absorption cross-section is obtained from \citet{Laor&Draine93}.
The dust moves by the radiation with the gas in the fixed dust-to-gas ratio, but exceptionally, the dust is assumed to be sublimated and sputtered if the dust temperature $T_\mathrm{dust}>1500$\,K and the gas temperature $T_\mathrm{gas}>10^5$\,K, respectively.
In contrast to the UV flux, the X-ray flux from the corona of the accretion disk is assumed to be isotropic for simplicity \citep{Netzer+87,Xu15}.
Cooling functions for $20\,\text{K} \leqslant T_\mathrm{gas} \leqslant 10^8\,\text{K}$ are used \citep{Meijerink&Spaans05,Wada+09} assuming solar metallicity.

The non-equilibrium chemistry for X-ray-dominated regions (XDRs) is also solved to reveal the multiphase and multilayered nature of the AGN environment.
From the chemical network of \citet{Meijerink&Spaans05} and \citet{Adamkovic+11}, 25 species are selected: H, H$_2$, H$^+$, H$_2^+$, H$_3^+$, H$^-$, e$^-$, O, O$_2$, O$^+$, O$_2^+$, O$_2$H$^+$, OH, OH$^+$, H$_2$O, H$_2$O$^+$, H$_3$O$^+$, C , C$^+$, CO, HCO$^+$, He, He$^+$, Na, and Na$^+$.
The reactions involving these species are solved for each grid cell per time step using the gas density, gas/dust temperature, and the ionization parameter.
The resulting abundances of the species are then advected based on the gas velocity for the simulation of the next time step.\footnote{To reduce the required computational memory, the advection is only tracked for H, H$^+$, H$_2$, O, O$^+$, H$_2$O, OH, C, C$^+$, and e$^-$.}

\begin{figure*}[tp]
\plotone{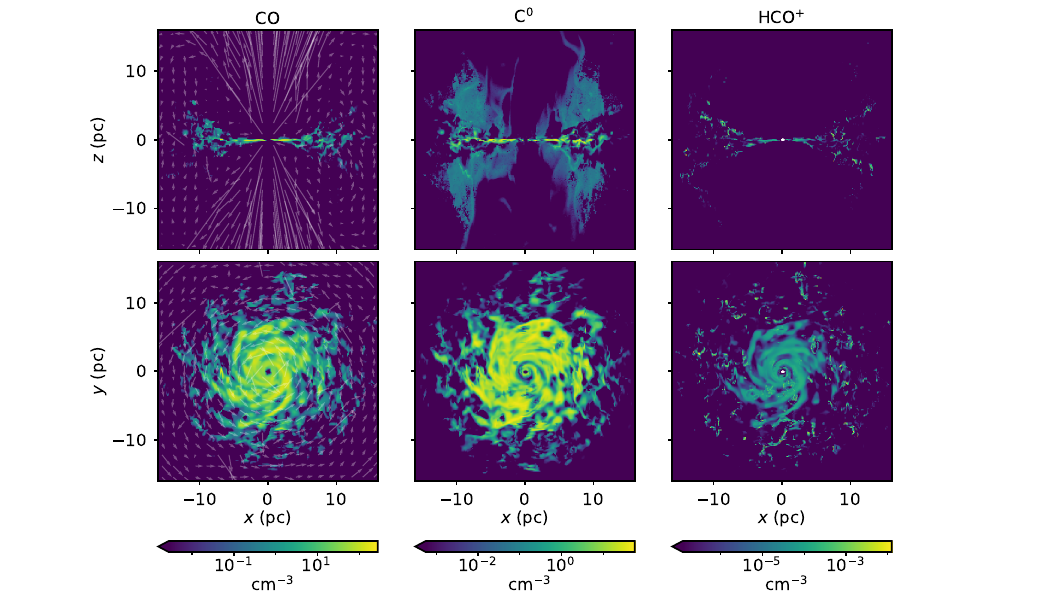}
\caption{
Comparison of the distributions of CO (left), C$^0$ (middle), and HCO$^+$ (right) on an axial plane ($x$-$z$ plane; top) and the equatorial plane ($x$-$y$ plane; bottom).
The arrows on the left panels represent the gas velocity field, which is common to all chemical species in the model.
\label{fig:denities}}
\end{figure*}

The snapshot used in this study is obtained when the model is in a quasi-steady state.
Figure \ref{fig:denities} presents the distributions of CO, C$^0$ (neutral carbon), and HCO$^+$.
The distributions of CO and C$^0$ are similar to those of H$_2$ and H$^0$, respectively \citepalias{W16}, where CO forms a thin disk, and C$^0$ has higher scale heights.
In contrast, HCO$^+$ is concentrated in the inner part ($r\lesssim5$\,pc) of the disk and is inhomogeneously distributed in the outer part of the disk plane and on a thin biconical surface corresponding to the boundary between the molecule-rich and ion-rich regions \citepalias{W16}.
This may be due to the reaction $\mathrm{CO + H_3^+ \longrightarrow HCO^+ + H_2}$.
In the following radiative transfer calculations, while the original snapshot consists of $256^3$ grid cells, we average it to $128^3$ cells to reduce the computational cost.

\subsection{Dust Continuum Radiative Transfer}

We compute the radiative transfer of the dust continuum based on the fountain model using the RADMC-3D code \citep[version 2.0;][]{RADMC-3D}, as done in the case of Paper \citetalias{P4} and \citetalias{M22}.
This code uses the dust distribution, grain species composition, and SED of the radiation source to calculate the temperature of each species in each cell, assuming radiative equilibrium.
This method takes into consideration the emission, absorption, and scattering caused by dust.

The dust distribution is derived from the gas distribution with a dust-to-gas ratio of 0.01 (Section \ref{sec:model}).
The dust grain composition is assumed to comprise four species, with two materials (silicate and graphite) and two sizes (0.01\,\micron\ and 0.1\,\micron).
The dust continuum intensity at submillimeter wavelengths can be adequately represented by the two bracketing grain sizes because the absorption cross-section is almost independent of the size and scattering is ineffective at wavelengths longer than 100\,\micron.
The mass ratio of the silicate and graphite grains is set as 0.625:0.375 \citep{Schartmann+05}.
The abundance ratio of the different grain sizes is determined according to the MRN distribution \citep{Mathis+77}, where the grain number density is proportional to $a^{-3.5}$, and $a$ represents the grain radius.
The absorption and scattering efficiencies of each grain species are obtained from \citet{Laor&Draine93}.

Only the central AGN is included as a radiation source for the heated dust.
The SED form of \citet{Schartmann+05} is used over a wavelength range of $10^{-3}$ to $10^4$\,\micron, with normalization to the bolometric luminosity of $5\times10^{43}$\,erg\,\per{s} (Section \ref{sec:model}).
However, for the sake of simplicity, radiation anisotropy is not assumed.

\subsection{Non-LTE Gas Line Radiative Transfer}

We solve the non-LTE radiative transfer of gas lines using the same method as that used in previous works (\citealt{Wada&Tomisaka05}; \citealt{Yamada+07}; \citetalias{M22}; Papers \citetalias{P1}, \citetalias{P2}, and \citetalias{P4}), which are based on the Monte Carlo and long-characteristic code of \citet{Hogerheijde&Tak00}.
The treatment of the dust continuum is generally identical to that in Paper \citetalias{P4} and \citetalias{M22}.
% In this study, we have refactored the Fortran codes used in the previous works with a more modern syntax.
In another method, \citet{Matsumoto+23} began line calculations of the fountain model using the versatile radiative transfer code, SKIRT.
Our method has an advantage over SKIRT in that we can examine how the line intensity changes along the line of sight (LOS).

Our code determines the non-LTE level population of the gas of interest using the following iterative procedure.
\begin{enumerate}
    \item Set LTE as the initial population for all the grid cells.
    \item Generate a large number of rays with random directions for each cell.
    \item Calculate the radiative transfer about a gas line
        \begin{equation}
            \frac{d I_\nu}{d\tau_\nu} = -I_\nu + S_\nu
        \end{equation}
        for each ray from the grid boundary to the cell wall.
        Here, the total (gas+dust) optical depth $\tau_\nu$ and source function $S_\nu$ depend on the gas level population $n_i$ in a cell through which the ray passes, as
        \begin{eqnarray}
            d\tau_\nu &=& \Big(\alpha_{\nu,\mathrm{gas}} + \sum\nolimits_k \alpha_{\nu,k}\Big) ds, \\
            S_\nu &=& \frac{j_{\nu,\mathrm{gas}} + \sum_k \alpha_{\nu,k}B_\nu(T_k)}{\alpha_{\nu,\mathrm{gas}} + \sum_k \alpha_{\nu,k}}, \\
            \alpha_{\nu,\mathrm{gas}} &=& \frac{h\nu_0}{4\pi} (n_l B_{lu} - n_u B_{ul}) \phi(\nu), \\
            j_{\nu,\mathrm{gas}} &=& \frac{h\nu_0}{4\pi} n_u A_{ul} \phi(\nu),
        \end{eqnarray}
        where $\nu_0$ is the line frequency, $\phi(\nu)$ is the line profile function, and $k$ is a label for dust grain species.
        We adopt a Gaussian line profile that is broadened owing to the thermal velocity and microturbulence, as
        \begin{eqnarray}
            \phi(\nu) &=& \frac{1}{\sqrt{\pi}\Delta\nu} \exp\left[-\frac{(\nu-\nu_0)^2}{\Delta\nu^2}\right], \\
            \Delta\nu &=& \nu_0 \frac{\sqrt{2k_\mathrm{B}T_\mathrm{g}/m+v_\mathrm{turb}^2}}{c},
        \end{eqnarray}
        where $T_\mathrm{g}$ is the gas temperature, $m$ is the mass of the gas molecule or atom, and $v_\mathrm{turb}$ is the microturbulence within a cell.\footnote{Here $v_\mathrm{turb}$ is defined as the most probable speed in three dimensions \citep{Hogerheijde&Tak00}. The standard deviation in one dimension such as LOS is $v_\mathrm{turb}/\sqrt{2}$.}
    \item Simultaneously solve both the radiative transfer from the cell wall to the cell center and the statistical equilibrium
        \begin{eqnarray}
            \frac{dn_i}{dt} &=& \sum_{j>i}n_j A_{ji} + \sum_{j\ne i}n_j(B_{ji}J_\nu + C_{ji}) \nonumber \\
            &-& n_i \Big[\sum_{j<i}A_{ij} + \sum_{j\ne i}(B_{ij}J_\nu + C_{ij})\Big] = 0,
        \end{eqnarray}
        where $J_\nu$ is the mean intensity at the cell center and is obtained by averaging the final $I_\nu$ of all rays, and $A_{ij}$, $B_{ij}$, and $C_{ij}$ are Einstein coefficients and collision coefficients.
        Since $J_\nu$ and $n_i$ are mutually dependent owing to the transfer from the cell wall, this step is iterated several times for convergence.
    \item After updating the level population for all the cells, iterate steps 3 and 4 until the result converges.
\end{enumerate}

For CO and HCO$^+$, we include rotational states up to $J$=15 in the calculations; thus, 15 lines from $J$=1--0 to $J$=15--14 are considered according to the selection rule.
For C$^0$, the $^3P_0$, $^3P_0$, and $^3P_0$ states are included, and all transitions [\ion{C}{1}](1--0), (2--1), and (2--0) are considered.
The energy levels and Einstein coefficients of CO, C$^0$, and HCO$^+$ are obtained from the Leiden Atomic and Molecular Database \citep[LAMDA;][]{LAMDA}.
For the collision rate coefficients with various partners, those available from LAMDA are used as a function of the gas temperature.
For CO and HCO$^+$, collisions with H$_2$ are considered, while for C$^0$, those with H$_2$, H$^0$, H$^+$, and e$^-$ are considered.
In cases wherein the collision rate coefficients are provided separately for ortho- and para-H$_2$, a constant ortho-to-para ratio of 3 is assumed.
A total of 1,000 Monte Carlo rays are generated per cell.
The microturbulence $v_\mathrm{turb}$ is set as 14.1\,\kmps, resulting in a LOS velocity dispersion of 10\,\kmps.\footnote{This $v_\mathrm{turb}$ is an intermediate value among those used in this series of papers. Consistent with this, the velocity dispersion of the HCN gas in the innermost 2\,pc of the Circinus galaxy is estimated to be $\sim$10\,\kmps\ at a spatial resolution of 0.5\,pc \citepalias{I23}.}
The external radiation incident on the grid boundary is assumed to be the cosmic microwave background ($T_\mathrm{bg}=2.73$\,K).
The iteration (step 5 in the previous paragraph) is repeated until the level population converges to a relative change of less than $10^{-3}$ on average, across all the cells.

After determining the level population, the model grid is pseudo-observed from any viewing direction or LOS.
In this paper, the inclination ($i$) of the LOS is measured from the $+z$ direction of the model, and the azimuthal angle ($\varphi$) of the LOS $xy$ projection is measured clockwise from the $-y$ direction.
The radiative transfer equation is integrated along the LOS, spanning from the far-side intersection with the grid boundary to the near-side intersection, to create a spectral cube with two spatial dimensions and one spectral dimension.
The spectral axis is sampled over the LOS velocity (\vlos) range of $\pm$125\,\kmps\ at 5\,\kmps\ intervals.
Thus, there are 51 velocity channels.
The sign of \vlos\ is taken such that positive values correspond to redshifts.

\section{Results} \label{sec:results}

\begin{figure*}[tp]
\plotone{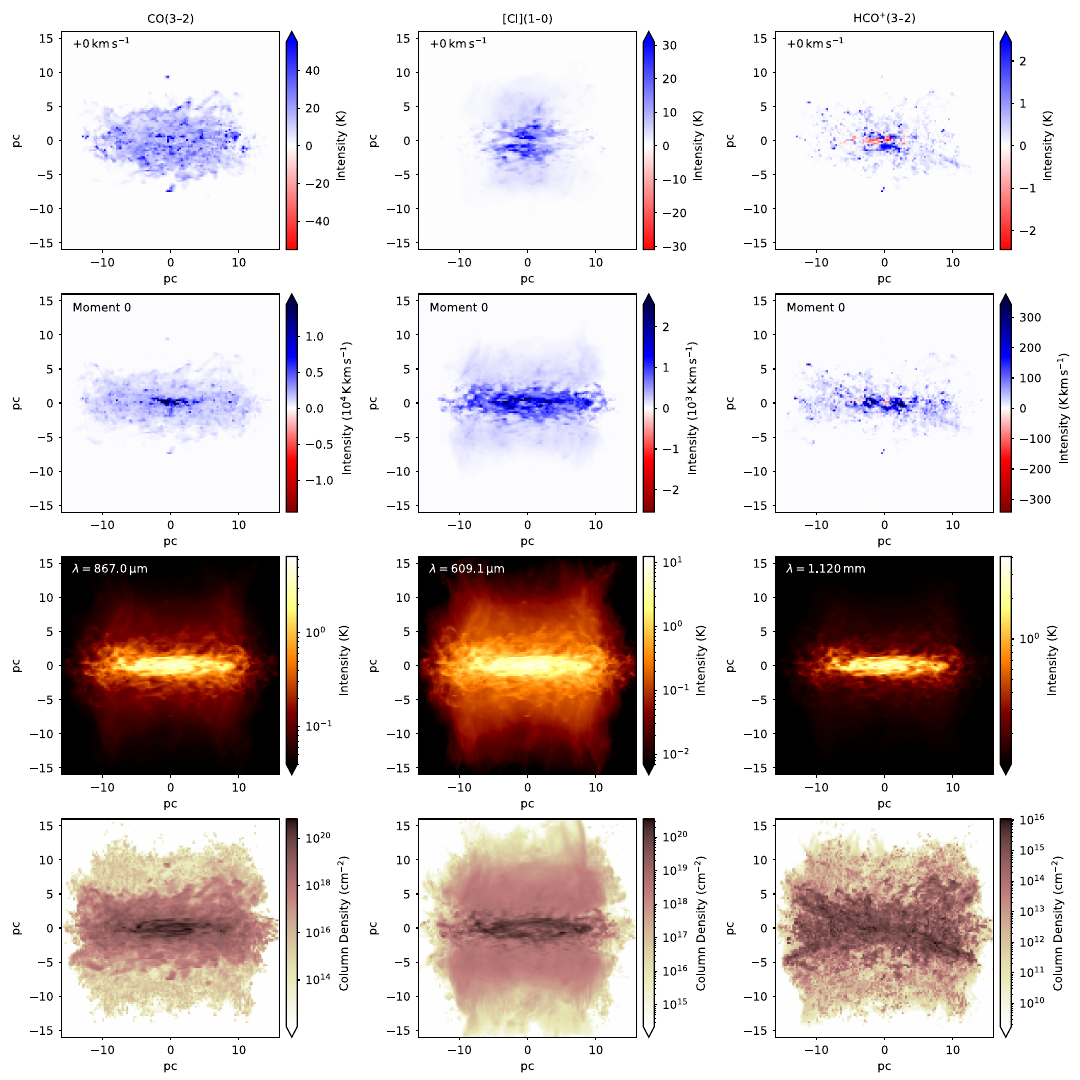}
\caption{
CO(3--2), [\ion{C}{1}](1--0), and HCO$^+$(3--2) lines pseudo-observed from the LOS direction of $(i,\varphi)=(80^\circ,0^\circ)$ (left, middle, and right columns, respectively).
Top row: Line center channel maps ($v_\mathrm{LOS}=0\,\kmps$).
Second row: Integrated intensity (moment 0) maps.
Third row: Underlying dust continuum maps.
Bottom row: CO, C$^0$, and HCO$^+$ column densities.
\label{fig:images}}
\end{figure*}

Figure \ref{fig:images} presents the results of the pseudo-observations of the CO(3--2), [\ion{C}{1}](1--0), and HCO$^+$(3--2) lines in the LOS direction $(i,\varphi)=(80^\circ,0^\circ)$ as an example.\footnote{We express the strength of radiation in terms of brightness temperature $T_\mathrm{b}$, which is related to intensity $I_\nu$ as $T_\mathrm{b}=\frac{c^2}{2k\nu^2}I_\nu$.}
In the line center channel maps ($\vlos=0\,\kmps$, top row of Figure \ref{fig:images}), CO(3--2) is horizontally elongated, whereas [\ion{C}{1}](1--0) extends in the polar direction.
This was also found in Paper \citetalias{P2}, wherein the dust continuum was ignored.
HCO$^+$(3--2) (267.558\,GHz), in contrast to the former two, presents a clumpy morphology and is observed as continuum absorption (negative line intensity) near the disk plane.
In Paper \citetalias{P4}, it was reported that CO(4--3) or higher transitions exhibited continuum absorption when the viewing angle was $85^\circ$ or greater; this is also the case in our results.
The HCO$^+$(3--2) continuum absorption is also visible in the velocity-integrated intensity map (moment zero, second row), although it is faint.
The absorption strength at the line center ($\sim$$-$2\,K) is approximately half the continuum level ($\sim$4\,K; third row).
The uniqueness of HCO$^+$ compared with the other two is also highlighted in the LOS column densities (bottom row).
In contrast to CO and C$^0$, HCO$^+$ is distributed not only in the disk plane, but also in the boundary between the molecule-rich and ion-rich regions above the disk plane (Section \ref{sec:model}), forming the ``X-shaped'' pattern.

We compare the observational features of the Circinus galaxy with the pseudo-observations.
While only the important PVDs are presented in the following subsections, all the major- and minor-axis PVDs of the CO, [\ion{C}{1}], and HCO$^+$ lines are presented in Appendix \ref{sec:allPVDs}.

\subsection{CO Major Axis PVD} \label{sec:COPVD}

\begin{figure*}[tp]
\plotone{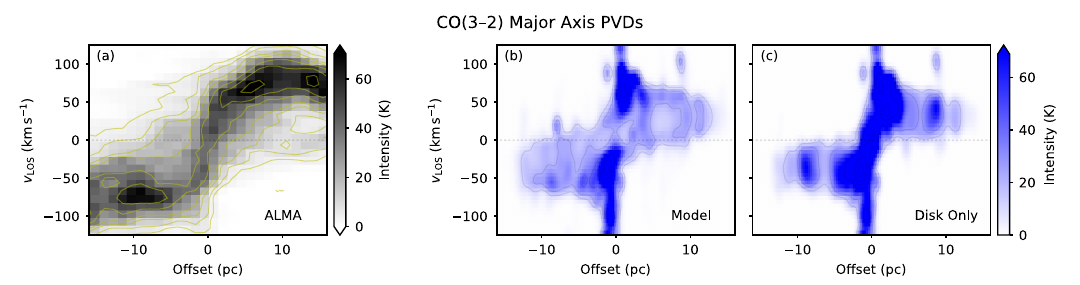}
\caption{
(a) CO(3--2) major axis PVD of the Circinus galaxy obtained with ALMA \citepalias{I23}.
The contours are drawn at 0.05, 0.1, 0.2, 0.3, 0.5, 0.7, and 0.9 times the maximum.
The horizontal dotted line indicates the systemic velocity.
(b) CO(3--2) major axis PVD of the fountain model pseudo-observed from $(i,\varphi)=(80^\circ,0^\circ)$.
The contours are drawn at the same fractional levels as in (a) relative to the maximum.
The slit width used to measure the average intensity at each offset and velocity is 1\,pc.
The diagram is blurred along the slit with a Gaussian kernel of 1\,pc FWHM.
(c) Same as (b), but created using only the contribution from the 0.5\,pc thick disk plane of the model.
\label{fig:CO_PVDs}}
\end{figure*}

Figure \ref{fig:CO_PVDs}(a) shows the CO(3--2) major axis PVD of the Circinus galaxy obtained using ALMA \citepalias{I23}.
The relatively high intensity ($\gtrsim$40\,K) reflects the rotation of the circumnuclear disk.
There is also a faint component that is close to the systemic velocity at $r>6$\,pc, which \citetalias{I23} attributed to a molecular inflow moving within the disk plane.

On the other hand, Figure \ref{fig:CO_PVDs}(b) presents the PVD of the fountain model pseudo-observed from $(i,\varphi)=(80^\circ,0^\circ)$.
This inclination is consistent with that of the Circinus galaxy constrained by  \citetalias{I23} through 3D dynamical modeling of the line.
The rotational component in the model is visible as relatively bright edges at $\pm$55\,\kmps.
This velocity is slightly smaller than the observed one ($\sim$70\,\kmps), and the pattern is not as obvious as in the observation.
This difference may be due to a lack of fine-tuning of the gravitational potential in the model, or to the fact that outer gas structures ($r>16$\,pc) were also observed with ALMA in the foreground and background.
The model PVD reproduces the faint slow component as found in the ALMA.
%, suggesting that this component is at least a realistic feature of the gas around AGN, rather than a feature of the host galaxy or an artifact.
It lies on the systemic velocity even at off-nuclear positions.
The origin of this component is investigated below.
The model also shows high-velocity components near the nucleus, while the observed PVD is ambiguous regarding the existence of such a component, although it is only implied on the blue-shifted side.
The vertical pattern around the systemic velocity at the nucleus found in the observation is less clear in the model.
This is due to self-absorption by the foreground gas (see below).
The possibility of this effect was also pointed out in the observation \citepalias{I23}.

To confirm whether the slow component originates from the disk plane, as interpreted in the observation, we create a PVD by extracting only the contribution from the disk plane of 0.5\,pc thickness (Figure \ref{fig:CO_PVDs}(c)).
The results show that the emission almost disappears at $\vlos=0\,\kmps$.
Thus, the slow component primarily originates from a region other than the disk plane.
Figure \ref{fig:CO_PVDs}(c) also shows a higher intensity than \ref{fig:CO_PVDs}(b) at the nucleus, indicating that the disk emission in this region is self-absorbed.

\begin{figure}[tp]
\plotone{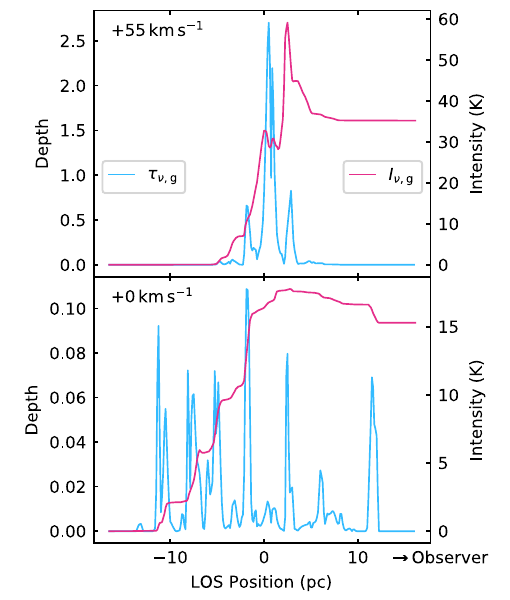}
\caption{
Growth curves of the rotational and slow components in the CO(3--2) model PVD (Figure \ref{fig:CO_PVDs}(b)).
The CO(3--2) intensity ($I_{\nu,\mathrm{g}}$) and optical depth ($\tau_{\nu,\mathrm{g}}$) without dust contributions, at offset $+$6\,pc and LOS velocity $+$55\,\kmps\ (rotational component, top) and 0\,\kmps\ (slow component, bottom) in the PVD, are presented as a function of the position on the LOS.
The origin of the position corresponds to the intersection with the equatorial plane of the model.
The observer is at $+\infty$.
The final line intensities observed from outside the model are the rightmost values, which are 35\,K and 16\,K at $+$55\,\kmps\ and 0\,\kmps, respectively.
\label{fig:CO_growth}}
\end{figure}

Figure \ref{fig:CO_growth} shows how the intensity of the rotational component ($+$55\,\kmps) and slow component (0\,\kmps) of the full-model PVD (Figure \ref{fig:CO_PVDs}(b)) at an offset $+$6\,pc increases along the LOS.
% Here, the origin of the position on the LOS is set such that it and the center of the model are equidistant from the observer at infinity.
Here, the observer is at infinity, and the origin of the position on the LOS is defined by the distance from the observer to the center of the model.
In this case, the origin corresponds to the intersection with the equatorial plane.
The intensity of the rotational component increases locally at the disk plane and then decreases by $\sim$40\% owing to self-absorption.
In contrast, the intensity of the slow component increases intermittently over a wide range along the LOS.
The range is approximately from $-$11\,pc to 0\,pc on the LOS, which corresponds to a region with a thickness of $11\,\mathrm{pc}\times \cos i = 2\,\mathrm{pc}$ just below the equatorial plane in the model.
This indicates that the slow component visible in Figure \ref{fig:CO_PVDs}(b) originates from a region slightly above or below the equatorial plane.
The structure and principle of the slow CO component are detailed in Section \ref{sec:slowCO}.

\subsection{[\ion{C}{1}] Minor Axis PVD} \label{sec:CIPVD}

\begin{figure*}[tp]
\plotone{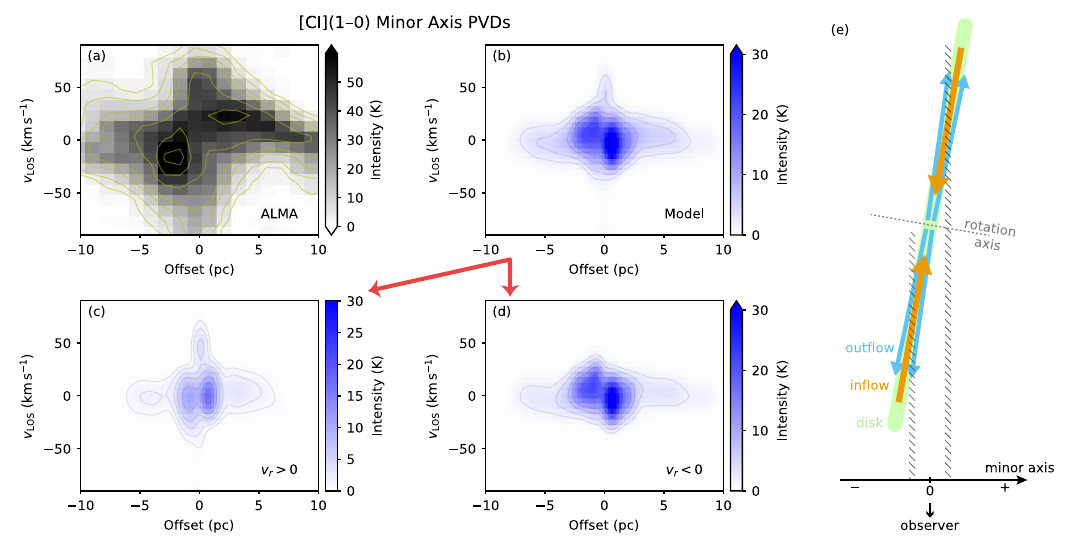}
\caption{
(a) [\ion{C}{1}](1--0) minor axis PVD of the Circinus galaxy obtained with ALMA \citepalias{I23}.
The contours are drawn at 0.05, 0.1, 0.2, 0.3, 0.5, 0.7, and 0.9 times the maximum.
(b) [\ion{C}{1}](1--0) minor axis PVD of the fountain model pseudo-observed from $(i,\varphi)=(80^\circ,0^\circ)$.
The contours are drawn at the same fractional levels as in (a) relative to the maximum.
The slit width used to measure the average intensity at each offset and velocity is 1\,pc.
The diagram is also blurred along the slit with a Gaussian kernel of 1\,pc FWHM.
(c) and (d) Same as (b), but created using only the contributions from grid cells where the gas radial velocity $v_r$ is positive and negative, respectively.
The color scales are matched.
(e) Schematic of which parts of the disk are observed at positive and negative offsets on the minor axis.
The hatched regions indicate the LOSs where the offset peaks are observed.
\label{fig:CI_PVDs}}
\end{figure*}

The minor-axis PVD of [\ion{C}{1}](1--0) observed in the Circinus galaxy \citepalias{I23} is shown in Figure \ref{fig:CI_PVDs}(a).
The axis direction is defined such that the far side of the disk (the side with the rotation axis inclined toward the observer) is positive (see Figure \ref{fig:CI_PVDs}(e)).
There are blueshifted and redshifted bright peaks at the negative and positive offsets, respectively, which \citetalias{I23} attributes to outflows.
Figure \ref{fig:CI_PVDs}(b) shows the model PVD pseudo-observed from $(i,\varphi)=(80^\circ,0^\circ)$.
Again, there are two offset peaks, but their velocity shifts are opposite to the observation: the redshifted one is at the negative offset.
Thus, the model does not reproduce the pattern observed in the Circinus galaxy.

To isolate the latent contributions of the outflows in the model, we classify the grid cells according to the sign of the radial component of the gas velocity ($v_r$) and extract only their respective contributions to create their PVDs separately (Figures \ref{fig:CI_PVDs}(c) and (d)).
The PVD of the negative $v_r$ cells (5d) is almost identical to that of the entire model (5b), thus revealing the dominance of these cells.
In contrast, the PVD of the positive $v_r$ cells (i.e., outflows; 5c) is qualitatively similar to the observed pattern of the two peaks (5a).
Therefore, we suspect that the observed PVD originates from two distinct gas flows.
The origins of the peaks in the positive and negative $v_r$ PVDs are the outflows and inflows near the disk plane, respectively.
This geometry is presented schematically in Figure \ref{fig:CI_PVDs}(e).
These results support the interpretation that the offset peaks in the Circinus galaxy are due to outflows, but also suggest that such outflows are overwhelmed by inflows in the current fountain model.
How this difference might be resolved is discussed in Section \ref{sec:CIpeaks}.
Figure \ref{fig:CI_PVDs}(c) also suggests that the high-velocity component near the nucleus in the observed and entire-model PVDs originates from an outflow.

\subsection{HCO$^+$ Line Profiles} \label{sec:profiles}

\begin{figure*}[tp]
\plotone{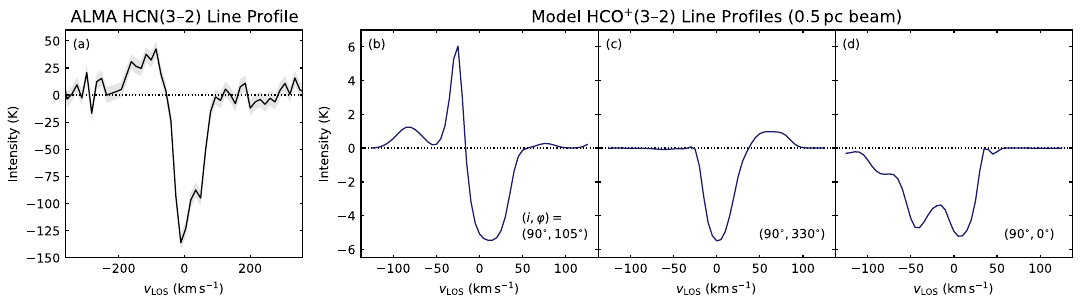}
\caption{
(a) Inverse P-Cygni profile of the HCN(3--2) line observed with ALMA at the nucleus of the Circinus galaxy \citepalias{I23}.
The beam size is $0.6\,\mathrm{pc}\times0.5\,\mathrm{pc}$.
(b) Example of an inverse P-Cygni profile of the HCO$^+$(3--2) line pseudo-observed at the center of the fountain model in the edge-on view using a 0.5\,pc beam.
(c) and (d) Same as (b), but from different azimuthal directions, resulting in a normal P-Cygni profile and a complicated absorption profile, respectively.
\label{fig:profile_examples}}
\end{figure*}

\begin{figure*}[tp]
% \plotone{HCO+_J02_0.5pc_profile_comparison.pdf}
\centering
\includegraphics[scale=1.00]{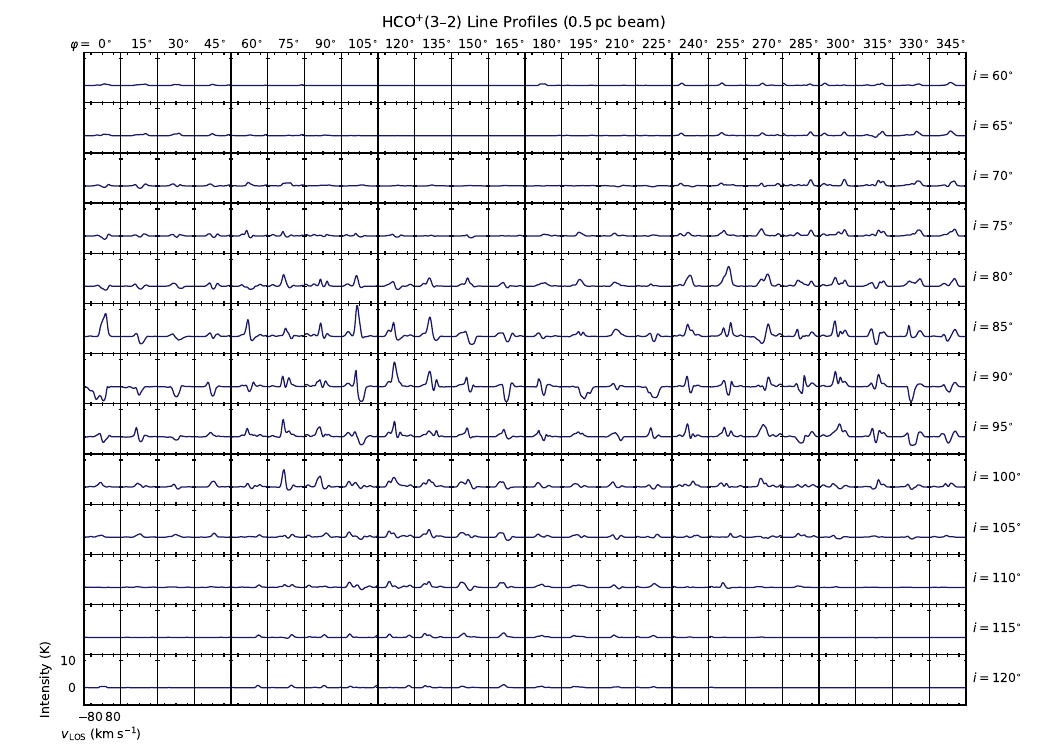}
\caption{
HCO$^+$(3--2) line profiles in the central 0.5\,pc beam pseudo-observed from various LOS orientations.
The rows and columns correspond to different inclination and azimuth angles, respectively.
The horizontal axis ticks are drawn at $\pm$80 and 0\,\kmps.
\label{fig:profiles}}
\end{figure*}

The inverse P-Cygni profile of the dense gas tracer HCN(3--2) detected with ALMA in the nucleus of the Circinus galaxy \citepalias{I23} is shown in Figure \ref{fig:profile_examples}(a).
This is considered strong evidence of an inflow.
Similarly, in the pseudo-observations of a similar dense gas tracer HCO$^+$(3--2) in the fountain model, inverse P-Cygni profiles can be obtained near the edge-on view with a beam size comparable to that of the ALMA observation (0.5\,pc).
A clear example is that from $(i,\varphi)=(90^\circ,105^\circ)$, as shown in Figure \ref{fig:profile_examples}(b).
Although the line shape is not very close to the observed one, the inverse P-Cygni nature, i.e. blueshifted emission and redshifted absorption, is evident.

However, the line profile has azimuthal dependence, and a normal P-Cygni profile or a complicated absorption profile can also be detected, even in the same edge-on view (Figures \ref{fig:profile_examples}(c) and (d)).
The LOS dependence of the HCO$^+$(3--2) line profile is summarized in Figure \ref{fig:profiles}, where the nuclear 0.5\,pc beam profile is presented for $i=60^\circ, 65^\circ, \cdots, 120^\circ$ and $\varphi=0^\circ, 15^\circ, \cdots, 345^\circ$.
The line intensity is higher when the LOS is closer to the edge-on view, especially at $\varphi\sim105^\circ$ (see also Figure \ref{fig:profile_examples}(b)) and at the opposite directions of $\varphi\sim285^\circ$.
It is also shown that the line profile varies significantly with the azimuth, even for the same inclination.
Weak continuum absorption is detected even when the inclination deviates $25^\circ$ from the edge-on view, but strong absorption predominating over emission is observed mainly within $\Delta i\sim5^\circ$ from the edge-on view.
This inclination range does not contradict the estimate derived from the ALMA observations \citepalias[$80^\circ$--$85^\circ$;][]{I23}.
However, it should be noted that in some cases, the line can be observed only in the emission, even in the exact edge-on view.

\begin{figure}[tp]
\plotone{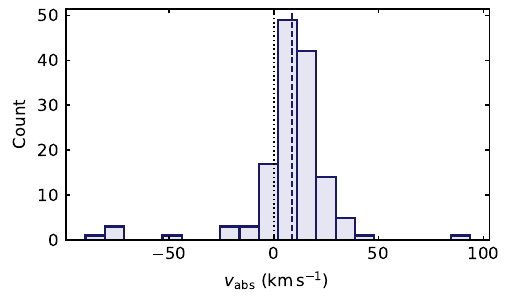}
\caption{
Histogram of the absorption velocity \vabs\ of the HCO$^+$(3--2) line profiles in the central 0.5\,pc beam (Figure \ref{fig:profiles}).
Profiles for which \vabs\ cannot be defined (no continuum absorption) are not included.
The vertical dashed line indicates the median of the distribution (8.7\,\kmps).
\label{fig:vabs}}
\end{figure}

To investigate what gas flows are observed in absorption, we measure the centroid of the absorption portion of each line profile in Figure \ref{fig:profiles} as
\begin{equation}
    \vabs \equiv \frac{\int_\mathrm{abs} v(-I_{\nu,\mathrm{line}})\,dv}{\int_\mathrm{abs} (-I_{\nu,\mathrm{line}})\,dv}, \label{eq:vabs}
\end{equation}
where the integral is applied only to the continuum absorption range.
The resulting \vabs\ values are shown in a histogram in Figure \ref{fig:vabs}, indicating that \vabs\ is systematically redshifted.
Hence, the continuum absorption of HCO$^+$(3--2) preferentially probes an inflow.
The median of the \vabs\ distribution is 8.7\,\kmps, which is close to the \vabs\ of the observed HCN(3--2) line profile (12\,\kmps; Figure \ref{fig:profile_examples}(a)).

\begin{figure*}[tp]
\plotone{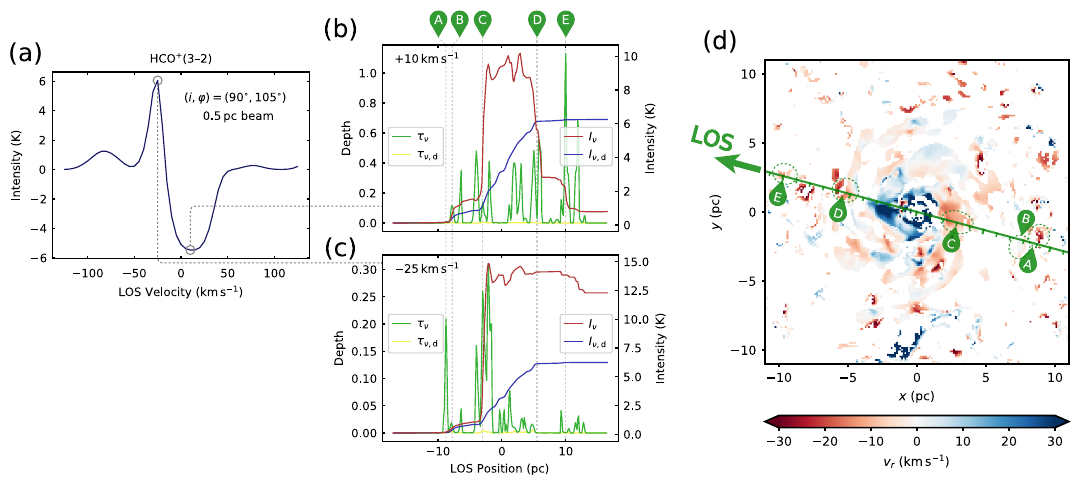}
\caption{
Example of the inverse P-Cygni profile of HCO$^+$(3--2) and its origin in the model.
(a) Inverse P-Cygni profile obtained from the LOS direction $(i,\varphi)=(90^\circ,105^\circ)$ with the nuclear 0.5\,pc beam (identical to Figure \ref{fig:profile_examples}(b)).
The absorption and emission peaks are at $+$10 and $-$25\,\kmps, respectively.
(b) and (c) Growth along the LOS of the total (gas$+$dust) intensity ($I_\nu$), dust continuum intensity ($I_{\nu,\mathrm{d}}$), and total and dust optical depths ($\tau_\nu,\tau_{\nu,\mathrm{d}}$) at $+$10 and $-$25\,\kmps, respectively.
The labels A to E indicate the coordinates at which significant changes occur (see text).
(d) Radial component of the gas velocity in the regions of high HCO$^+$ density ($n_\mathrm{HCO^+}>10^{-5}\,\percb{cm}$) within the disk plane of 0.5\,pc thickness.
The LOS coordinate is indicated by the solid green line with ticks in 1\,pc increments, and points A--E in panels (b) and (c) are also indicated.
An animation of the evolution of the line profile is presented in Figure \ref{fig:movie}.
\label{fig:origins}}
\end{figure*}

To identify the locations at which the inverse P-Cygni profile of HCO$^+$(3--2) originates, we examine the line intensity growth along the LOS, adopting $(i,\varphi)=(90^\circ,105^\circ)$ as an example.
Figure \ref{fig:origins}(a) shows the profile in this direction (identical to that in Figure \ref{fig:profile_examples}(b)).
The absorption and emission peaks are at $+$10 and $-$25\,\kmps, respectively.
Figures \ref{fig:origins}(b) and (c) present the total (gas $+$ dust) intensity, dust continuum intensity, and total and dust optical depths at the two velocity channels as functions of the position along the LOS.
The LOS coordinates of interest are labeled as A--E.
The locations of these points in the model are indicated in Figure \ref{fig:origins}(d), which also shows the gas radial velocity for the HCO$^+$-dense regions ($n_\mathrm{HCO^+}>10^{-5}\,\percb{cm}$) in the disk plane with a thickness of 0.5\,pc.
First, at location A on the far side of the disk, the line (total $-$ dust) intensity at $-$25\,\kmps\ is increased by a clump inflowing towards the nucleus, and immediately after that, at location B, the line intensity at $+$10\,\kmps\ is increased by a weakly outflowing clump.
In the cloud at location C, the line intensity is enhanced at both velocities because the cloud has a large velocity dispersion.
This cloud is the most significant contributor to the emission.
The dust intensity increases by a factor of $\sim$6 between locations C and D, and at location D, the line intensity at $+$10\,\kmps\ is diminished by two neighboring inflowing clumps, such that the total intensity becomes less than the dust continuum, thus causing continuum absorption.
The intensity also sharply decreases owing to the infalling (i.e., redshifted) clump located at E.
In contrast to Figure \ref{fig:origins}(b), Figure \ref{fig:origins}(c) shows that the $-$25\,\kmps\ intensity remains almost unchanged on the near side of the disk.
The evolution of the entire profile, including all the velocity channels, is presented in Appendix \ref{sec:movie}.

\section{Discussion} \label{sec:discussion}

\subsection{Origin of the Slow CO Component} \label{sec:slowCO}

In the central 20\,pc of the Circinus galaxy, \citetalias{I23} found a faint and slow emission component in the CO(3--2) major axis PVD close to the systemic velocity and interpreted it as a molecular inflow through the disk plane.
We have shown in Section \ref{sec:COPVD} that this slow component is reproduced by the fountain model, but its origin is mainly in the regions above and below the disk plane, in contrast to the case of the rotational component.
For the positive offset along the major axis (the side where the gas is rotating away from the observer), the origin is in the $\sim$2\,pc height region behind the disk.

\begin{figure}[tp]
\plotone{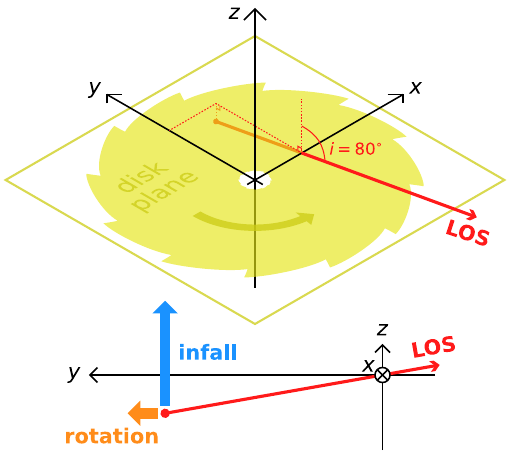}
\caption{
(Top) Direction of the LOS of $(i,\varphi)=(80^\circ, 0^\circ)$, which is the same as that in Figure \ref{fig:CO_PVDs}(b), at a positive offset along the kinematic major axis.
The disk is in the $xy$-plane and rotates counterclockwise.
The slow CO component originates behind the disk plane.
(Bottom) Side view of the LOS path.
Behind the disk, the rotation and infall velocities cancel each other out.
\label{fig:slow_CO}}
\end{figure}

The reason why the gas outside the disk appears to have a low LOS velocity must be that its rotational velocity tends to be canceled out by the falling motion to the disk plane, as shown in Figure \ref{fig:slow_CO}, even though the gas is turbulent owing to the supernovae.
The correspondence between the receding rotation side and the back of the disk is consistent with this interpretation.
The spikes in the bottom panel of Figure \ref{fig:CO_growth} likely represent clumps having LOS velocities close to zero after the fall and turbulence are combined.
We suggest that the slow component is a consequence of a three-dimensional structure that includes vertical motion, rather than a two-dimensional disk structure.
While inflows are present, their contribution to the line intensity is weaker than that of the failed winds, at least based on the current model.
It should be noted that the absolute rotational velocity of the main component in the current model ($\sim$55\,\kmps\ at $r=6$\,pc) is slightly less than the observed velocity ($\sim$70\,\kmps\ at the same radius), as shown in Figures \ref{fig:CO_PVDs}(a) and (b).
This difference could be resolved by adjusting the gravitational potential of the model.
Although such a modification could change the nature of the slow component, we do not pursue it in this paper.

It should be noted that the major-axis PVD of [\ion{C}{1}](1--0) lacks a similar slow component (see the upper-middle panel of Figure \ref{fig:allPVDs}).
This difference reflects the fact that the abundance of C$^0$ is small just above the disk plane, as shown in the upper-middle panel of Figure \ref{fig:denities}.
When observed at angles close to edge-on, the slow component does not appear because of the gap in the C$^0$ distribution. 
The ALMA observations of [\ion{C}{1}](1--0) also showed no slow components.
This consistency supports the multilayer structure predicted by the fountain model, in which the molecular and atomic phases exist at different scale heights.

\subsection{Implication from the [\ion{C}{1}] Offset Peaks} \label{sec:CIpeaks}

In Section \ref{sec:CIPVD}, we have shown that the outflow near the disk plane in the fountain model produces a sign of offset peaks in the [\ion{C}{1}](1--0) minor-axis PVD similar to those found in the Circinus galaxy.
However, the outflow component is overwhelmed by the inflow component in the model.
The former result supports the interpretation of \citetalias{I23} that the [\ion{C}{1}](1--0) offset peaks are due to outflows.
However, the latter result suggests that the outflow of the current model is weaker than that of the Circinus galaxy.
Consistent with this, the spatial offsets and velocity shifts of the peaks in the PVD of the model outflow component are smaller than those in the ALMA observations (Figure \ref{fig:CI_PVDs}).

The weakness of the outflows could be resolved by using a model snapshot at a time when the fountain is more active than in the current snapshot or by tuning the gravitational potential and/or AGN luminosity of the radiation-hydrodynamic simulation itself.
Actually, a recent estimate of $L_\mathrm{bol}$ for the Circinus galaxy based on X-ray observations is $1.3\times10^{44}\,\mathrm{erg\,\per{s}}$ \citep{Uematsu+21}, which is a factor of $\sim$2 higher than that adopted in the current model.
If a higher $L_\mathrm{bol}$ is adopted, the radiation-driven outflow could become faster and more massive, but it is not so straightforward how the pseudo-observations would change, since the XDR chemistry and the entire gas distribution structure, including the inflow and failed wind, would be affected at the same time.
Conversely, the pc-scale [\ion{C}{1}](1--0) offset peaks reflect the local structures inside the model and could be a good reference for exploring a more appropriate model setup to match real AGNs.

\subsection{Inflow Probed by HCO$^+$ Inverse P-Cygni Profile} \label{sec:inflow_rate}

One of the most important results of high-resolution (0.5\,pc) ALMA observations of the Circinus galaxy by \citetalias{I23} was the robust detection of a pc-scale molecular inflow into the AGN through the inverse P-Cygni profile of HCN(3--2).
In Section \ref{sec:profiles}, we have shown that, in the fountain model, the HCN-like dense gas tracer HCO$^+$ exhibits inverse P-Cygni profiles in $J$=3--2 at inclination angles consistent with the Circinus galaxy and that the absorption LOS velocities are on average close to that observed in the Circinus galaxy.
However, we have also found that the line profile varies greatly with the azimuth, resulting in normal P-Cygni, complicated absorption, and even pure emission profiles.

As shown in Figure \ref{fig:origins}, the HCO$^+$(3--2) inverse P-Cygni profile is caused by inflowing and outflowing clumps that happen to be on the LOS.
This inhomogeneous distribution is responsible for the azimuthal dependence of the line shape.
In addition, this geometry is far from the symmetric ones that are often assumed to explain (inverse) P-Cygni profiles in actual observations, such as plane-parallel gas layers and expanding shells.
Our results suggest that such geometry assumptions may be oversimplified for circumnuclear disks, at least for dense gas tracers.
It is also important to note that the clumps are far from the nucleus ($r>5$\,pc) compared to the beam size (0.5\,pc).
In general, it is impossible to determine the location of the absorbing gas in the LOS from an absorption line alone, and certain assumptions are required to be made.
For example, in \citetalias{I23}, a radius of half the beam size ($r=0.27$\,pc) was assumed to be a probabilistic expectation.
The absorber identified in the model is located more than an order of magnitude farther away than this assumption, thus making the effect of clumpiness on the angle dependence significant.

\begin{figure}[tp]
\plotone{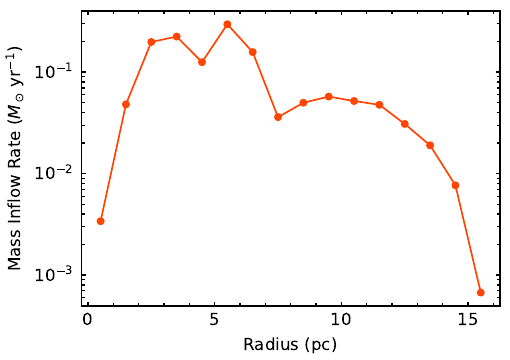}
\caption{
Mass inflow rate as a function of radius, azimuthally integrated within the disk plane of 0.5\,pc thickness.
\label{fig:inflow_rate}}
\end{figure}

The main motivation for observing the molecular inflow is to discuss the feeding of AGNs.
Then, what are the mass inflow rates carried by the clumps responsible for the absorption in the model?
By integrating the mass density of the clumps at locations D and E ($r=5.5$ and 10\,pc), we obtain $\sim$0.01\,$M_\odot$\,\per{yr} and $\sim$0.007\,$M_\odot$\,\per{yr}, respectively.
On the other hand, the azimuthally integrated mass inflow rate in the disk plane, as a function of radius, is presented in Figure \ref{fig:inflow_rate}.
The rate increases closer to the center, but decreases sharply at $r<2$\,pc, indicating that the majority of the gas is blown away as outflows.
In comparison with this figure, clump D (a pair of neighboring clumps) and clump E are found to be responsible for $\sim$4\% and $\sim$13\% of the total mass inflow rate at their radii, respectively.
Considering the size of the clumps ($\sim$1\,pc), the percentage of the inflowing mass accounted for by clump D does not deviate significantly from the circumferential average.
This is probably because clump D (region with a high HCO$^+$ abundance) is located in a spiral arm.
In contrast, Clump E occupies a much larger percentage than the other parts of the same radius.
This is because the distribution of H$_2$ itself is inhomogeneous at this radius.

The mass inflow rate estimated by \citetalias{I23} is 0.20--0.34\,$M_\odot$\,\per{yr}, which is almost consistent with the sum of the azimuthally integrated rates at the radii of clumps D and E in our model (0.35\,$M_\odot$\,\per{yr}).
However, this agreement is probably coincidental because the assumptions of the absorber location and uniform gas distribution made by the authors in their estimate are largely different from the situation in our model.
Our results suggest that there is scatter in the contribution of the HCO$^+$ absorbers to the total mass inflow rate.
The rate calculated from a single absorption profile could be overestimated by one order of magnitude.
This uncertainty should be taken into consideration in the future when a new inflow is detected as redshifted absorption.
It would be informative to check whether the obtained estimate of the total inflow rate is of the same order of magnitude as the rates suggested by the 3D hydrodynamic model (Figure \ref{fig:inflow_rate}).

\subsection{Detectability of Atomic Outflow}

In the fountain model, the fastest outflows occur at angles close to the rotation axis, but the ionized gas dominates in this region.
The pc-scale atomic outflows are slightly above the disk plane, at least in the current model.
The relatively diffuse hollow-cone atomic gas distribution at higher scale heights, represented by C$^0$ (Figure \ref{fig:denities}), is more like a secondary structure formed by the fast polar outflow than a direct outflow.
The inner boundary of the cone is outflowing owing to radiation, but the bulk region is composed of gas falling to the equatorial plane after being blown up.
Outflows near the disk plane can be observed as offset peaks in the minor-axis PVD of [\ion{C}{1}](1--0) (blue-/red-shifted on the near/far side of the disk, respectively) for a Type 2 AGN with outflows as strong as those of the Circinus galaxy, as described in Section \ref{sec:CIpeaks}.
Conversely, if offset peaks of the inverted velocity shifts are observed, as in our pseudo-observation, it suggests the presence of atomic inflows in the disk.

\begin{figure}[tp]
\plotone{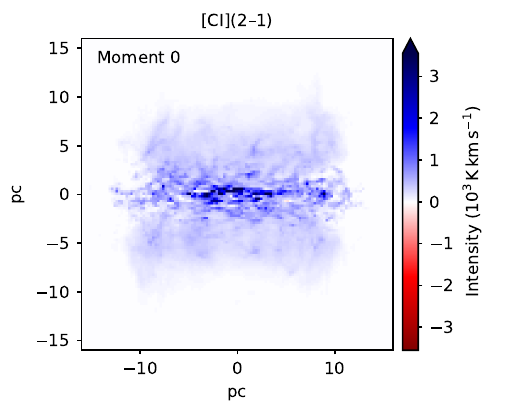}
\caption{
Moment 0 map of [\ion{C}{1}](2--1) created in the same manner as Figure \ref{fig:images}.
\label{fig:CI21mom0}}
\end{figure}

Because the gas is cold near the disk plane, [\ion{C}{1}](1--0), which has a low excitation energy, is suitable for observing the outflows around the disk.
Figure \ref{fig:CI21mom0} presents the moment 0 map for the higher energy-level line, [\ion{C}{1}]($^3P_1$--$^3P_0$) $\equiv$ [\ion{C}{1}](2--1).
By comparing this with the middle panel of the second row of Figure \ref{fig:images}, it can be found that the disk emission is more pronounced in [\ion{C}{1}](1--0), thus indicating its suitability for outflow (or inflow) detection.
In contrast, the polar elongated emission from the hollow cone distribution is more clearly imaged in [\ion{C}{1}](2--1) because the gas is hotter there.
Therefore, to observe the geometrically thick structure formed by the polar outflows, [\ion{C}{1}](2--1) is more suitable than [\ion{C}{1}](1--0).
Needless to say, however, actual observations must take into consideration practical issues such as poor atmospheric transmission in the high-frequency band for [\ion{C}{1}](2--1) (809.342\,GHz).

\subsection{Detectability of Molecular Inflows}

As an indication of molecular inflow, the slow component of CO emission in the major-axis PVD was suggested by \citetalias{I23}.
However, as discussed in Section \ref{sec:slowCO}, this component originates from the gas falling toward the disk.
This gas is not an inflow at this moment, although it eventually reaches the disk plane and becomes an inflow.
Therefore, molecular inflows would not be detectable in a CO major-axis PVD.
Based on the model, the inflow velocity of the gas in the disk is at most a few percent of the rotational velocity; thus, the inflow does not clearly deviate from the rotational component in the PVD.
Extracting the inflow from CO emissions requires three-dimensional kinematic modeling, including vertical motion.
Although the slow CO component is not an inflow, based on the fountain model, failed winds also have a significant importance in creating a quasi-steady gas circulation \citep{Wada12}.
Our results suggest that a pc-scale CO major-axis PVD allows us to capture the falling component, which has not been recognized observationally.

Clear evidence for molecular inflow is redshifted continuum absorption because it establishes that the gas is in front of the AGN (continuum source).
As shown in Section \ref{sec:profiles}, when the dense-gas tracer HCO$^+$(3--2) is observed with a beam size of 0.5\,pc, its nuclear spectrum shows weak continuum absorption even at a $25^\circ$ tilt from edge-on, and the absorption is systematically redshifted.
Figure \ref{fig:origins} shows that absorption does indeed occur in the inflowing clumps.
However, because of the clumpiness of the absorber, absorption is not always observed in any direction.
Because HCN would have a distribution similar to that of HCO$^+$, our results suggest that the detection of inflow using the dense-gas tracers is, to some extent, a probabilistic event.

\begin{figure*}[tp]
\plotone{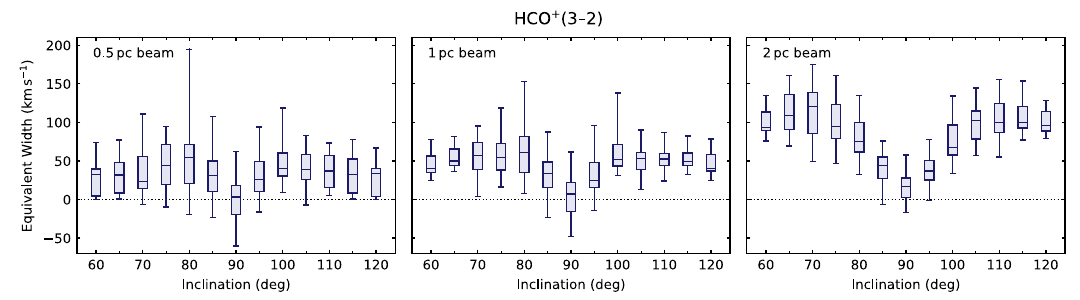}
\caption{
Left: Box plot of the EWs of the HCO$^+$(3--2) line profiles in the central 0.5\,pc beam (Figure \ref{fig:profiles}).
The distribution of the EWs obtained from the 24 azimuthal angles at each inclination is shown.
The lower and upper edges of the box indicate the 25th and 75th percentile points, the line inside the box indicates the median, and the caps of the whiskers indicate the maximum and minimum values.
Negative EWs correspond to profiles dominated by continuum absorption.
Middle and right: Same as the left panel, but for beam sizes of 1\,pc and 2\,pc, respectively.
\label{fig:EW_beams}}
\end{figure*}

To investigate how the probability of observing significant absorption depends on the LOS, we calculate the equivalent width (EW) of each line profile shown in Figure \ref{fig:profiles} obtained with the 0.5\,pc beam as
\begin{equation}
    \mathrm{EW} \equiv \frac{\int I_{\nu,\mathrm{line}}\,dv}{I_{\nu,\mathrm{cont}}}, \label{eq:EW}
\end{equation}
where $I_{\nu,\mathrm{line}}$ and $I_{\nu,\mathrm{cont}}$ represent the line and continuum intensities, respectively.
A negative EW indicates that the line is dominated by absorption.
The distribution of the resulting EWs for each inclination angle is shown as a box plot in Figure \ref{fig:EW_beams}.
It can be found that the fraction of absorption-dominated profiles reaches nearly 50\% only at $i=90^\circ$ and decreases to 10--20\% at $i=85, 95^\circ$.
This confirms the visual inspection of Figure \ref{fig:profiles} mentioned in Section \ref{sec:profiles} that strong absorption predominating over emission occurs only within $\pm$5$^\circ$ from the edge-on.

The probability of dominant absorption can be affected by the beam size to obtain a line profile, because if the beam is not filled by the continuum source, the emission from the gas not in front of the continuum source will cancel out the absorption.
To evaluate the beam size dependence, as done with the 0.5\,pc beam, we also obtain line profiles from the different LOS directions using 1\,pc and 2\,pc beams, calculate their EWs, and plot the resulting distributions (Figure \ref{fig:EW_beams} middle and right).
These panels show that the EW distribution moves positively with the beam size, as expected, owing to the effect of ambient emission.
Correspondingly, the occurrence of absorption-dominated profiles decreases, even at $i=90^\circ$, and the rate is less than 25\% for the 2\,pc beam.
In summary, Figure \ref{fig:EW_beams} suggests that to detect prominent continuum absorption in HCO$^+$(3--2) with a high probability, it is important that the AGN be close to edge-on within $\lesssim$5$^\circ$ and that the beam size is as small as $\lesssim$1\,pc.

\begin{figure}[tp]
\plotone{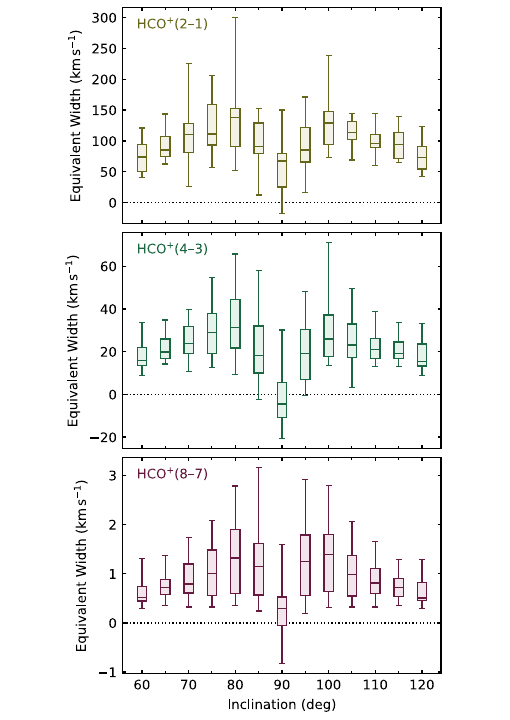}
\caption{
Box plot of the EWs of HCO$^+$(2--1), (4--3), and (8--7) at 1\,pc beam, created in the same manner as that in Figure \ref{fig:EW_beams}.
\label{fig:EW_levels}}
\end{figure}

The required proximity of the system to the edge-on view depends on the line excitation level.
We compare the inclination dependence of the EWs for the 1\,pc beam for HCO$^+$(2--1), (4--3), and (8--7) in Figure \ref{fig:EW_levels}.
This figure shows that the more highly excited the transition is, the stronger the absorption is in the exact edge-on case; however, the inclination dependence of the EWs simultaneously becomes steeper.
Therefore, the allowable range of inclinations for detecting an absorption-dominated profile becomes narrower.
At higher excitation levels (higher frequencies), the dust continuum becomes more pronounced than gas emission, and when observed, absorption becomes dominant.
However, the dense regions where HCO$^+$ can be excited to such high levels simultaneously become more concentrated in the equatorial plane.
The trade-off between the absorption intensity and inclination tolerance must be considered when selecting a line for detecting absorption, as well as the line strength and observation time.

In Paper \citetalias{P4}, we performed a similar test on the detectability of CO lines and found that continuum absorption in CO lines can be detected at CO(4--3) or higher levels when the LOS is within $5^\circ$ of the edge-on and the beam size is $\leq$1\,pc, and that if detected, the absorbers are located at $r=10$--15\,pc from the nucleus.
Compared with CO, HCO$^+$ lines probe absorbers closer to the center and have a wider inclination tolerance.
In addition, the HCO$^+$ lines exhibit more absorption-dominated profiles than the CO lines (see the figures in Paper \citetalias{P4}).
These points reinforce the strategy often adopted in observations that dense gas tracers such as HCO$^+$ and HCN are more suitable for detecting molecular inflow than CO, which traces relatively diffuse gases.

In the radiative transfer calculations of the submillimeter lines in this study and in Paper \citetalias{P4}, only the thermal emission from the dust heated by the AGN is assumed to be a source of continuum light.
However, \citet{Kawamuro+22} systematically analyzed ALMA data for 98 local hard X-ray-selected AGNs and found a tight correlation between the nuclear 1.3\,mm luminosity and 14--150\,keV luminosity.
This suggests that the millimeter emission originates near the X-ray corona.
The authors argue that dust emission from the torus would not be significant because the observed millimeter spectral slope is generally flatter than expected for thermal emission.
If there is a dominant compact continuum source other than dust, absorption is more likely to be detected, and the requirements for inclination and beam size can be relaxed.
If the origin of the continuum can be quantitatively understood from observations, more accurate predictions can be made by adding such a component to the model.

In addition to the submillimeter lines, absorption detectability with CO was also verified for ro-vibrational transitions in the infrared ($\sim$4.7\,\micron) by \citetalias{M22}.
The authors reported that the continuum source at this wavelength is a compact region of $r\sim1.5$\,pc in the innermost part of the gas distribution and that inflows and outflows at $r=1$--2\,pc can be detected as absorption components.
In the submillimeter HCO$^+$ lines calculated here, the continuum source is disk-shaped with $r\sim5$\,pc and absorption occurs at $r=5$--10\,pc (Figures \ref{fig:images},\ref{fig:origins}).
In comparison, the infrared CO lines probe further inward.
In addition, because the continuum source is point-like, and the region at which CO is vibrationally excited is small, the absorption intensity is almost independent of the beam size.
This is advantageous for detecting molecular inflow.
However, because of the self-shielding of dust radiation in the infrared wavelengths, absorption cannot be detected when the LOS is close to edge-on, and an intermediate inclination of $i=50$--$80^\circ$ is required \citepalias{M22}.
This is in contrast to the situation in the submillimeter lines, where the absorption is stronger closer to the edge-on.
Conversely, the infrared CO lines and submillimeter dense-gas lines can be complementary inflow tracers, depending on the inclination of the target AGN.

It should be noted that inflow detection using absorption lines is a microscopic event that reflects internal structures such as clumps.
The inhomogeneous structures of the circumnuclear disk depend on various parameters such as the SMBH mass, total gas mass, supernova rate, and Eddington ratio \citep{Wada&Norman02,Wada15}.
Conversely, this also implies that absorption observations can be used to statistically determine such model parameters if we obtain a sufficiently large statistical sample of submillimeter or infrared absorption lines in the Circinus galaxy and other nearby AGNs.

\section{Conclusions} \label{sec:conclusions}

Recent high-spatial-resolution ALMA observations have revealed signs of pc-scale AGN inflow and outflow \citepalias{I23}.
In this study, we have presented predictions that can be compared with such observations by solving the radiative transfer problem for 3D radiation-hydrodynamic calculations (the radiation-driven fountain model).
Our findings can be summarized as follows.

\begin{enumerate}
    \item The CO(3--2) major axis PVD has reproduced a slow, faint emission similar to the actual observations.
    However, this component does not originate from inflows moving inward in the disk plane, but from failed winds failing down to the disk plane.
    This result suggests that the interpretation of a PVD of CO, a medium-density gas tracer, must take into consideration three-dimensional kinematic structures, including vertical motions.
    \item The [\ion{C}{1}](1--0) minor axis PVD exhibits two offset peaks with velocity shifts opposite to those observed, but when created only from gas with positive radial velocities, it shows a sign of peaks shifted in the same directions as observed, suggesting that the observed peaks are due to outflows, but the model outflow may be too weak to explain the observation.
    The higher frequency line [\ion{C}{1}](2--1) has larger contributions from hot gas at high scale heights, thus helping us to observe the geometrically thick structure formed by outflows.
    \item The nuclear spectrum of HCO$^+$(3--2) can show an inverse P-Cygni profile similar to that observed for HCN(3--2) when the inclination angle is close to edge-on.
    However, the profile has an azimuthal dependence because the absorption is caused by inflowing clumps that are unevenly distributed at 5--10\,pc from the nucleus.
    Such clumps do not necessarily represent the total mass inflow rate well at their orbital radii.
    \item The probability of obtaining an absorption-dominated HCO$^+$(3--2) profile is a few 10\% for a beam size of 0.5\,pc and an inclination of $85^\circ$--$95^\circ$.
    This probability decreases as the beam size increases owing to the larger contribution of the ambient emission.
    For higher excited HCO$^+$ lines, the absorption becomes more significant when the LOS is exactly edge-on.
    However, the inclination tolerance for the detection of absorption is also reduced.
    \item Compared to CO rotational absorption \citepalias[Paper][]{P4}, HCO$^+$ absorption can probe more inner clumps over a wider range of inclination angles.
    In this respect, dense gas tracers are more suitable than CO in detecting inflows.
    In addition, HCO$^+$ absorption can be detected in an inclination range complementary to that of near-infrared CO ro-vibrational absorption \citepalias{M22}.
\end{enumerate}

Our results suggest that pc-scale observations of dense gas tracers such as HCN and HCO$^+$ and neutral [\ion{C}{1}] lines are effective for detecting AGN molecular inflow and atomic outflow, respectively.
We expect that conducting high-resolution observations and comparing the results with radiative transfer predictions will facilitate the building of a refined, realistic hydrodynamic model and the derivation of a concrete, reliable observational interpretation.

\begin{acknowledgments}
We thank the anonymous referee for helpful comments that improved the paper.
Numerical computations were carried out on Cray XC50, Small Parallel Computers, and analysis servers at Center for Computational Astrophysics, National Astronomical Observatory of Japan.
This work is supported by the Japan Society for the Promotion of Science (JSPS) KAKENHI Grant Numbers JP21H04496 (K.W., T.I., and S.B.) and JP20K14531 (T.I.).
K.M. is a Ph.D. fellow of the Flemish Fund for Scientific Research (FWO-Vlaanderen) and acknowledges the financial support provided through Grant number 1169822N.
\end{acknowledgments}

\vspace{5mm}
\software{RADMC-3D \citep{RADMC-3D},
NumPy \citep{NumPy},
SciPy \citep{SciPy},
Matplotlib \citep{Matplotlib},
Astropy \citep{Astropy_v5}}

\appendix

\section{All Major and Minor PVD\lowercase{s} of the CO, [\ion{C}{1}], and HCO$^+$ Lines} \label{sec:allPVDs}

\begin{figure*}[tp]
\plotone{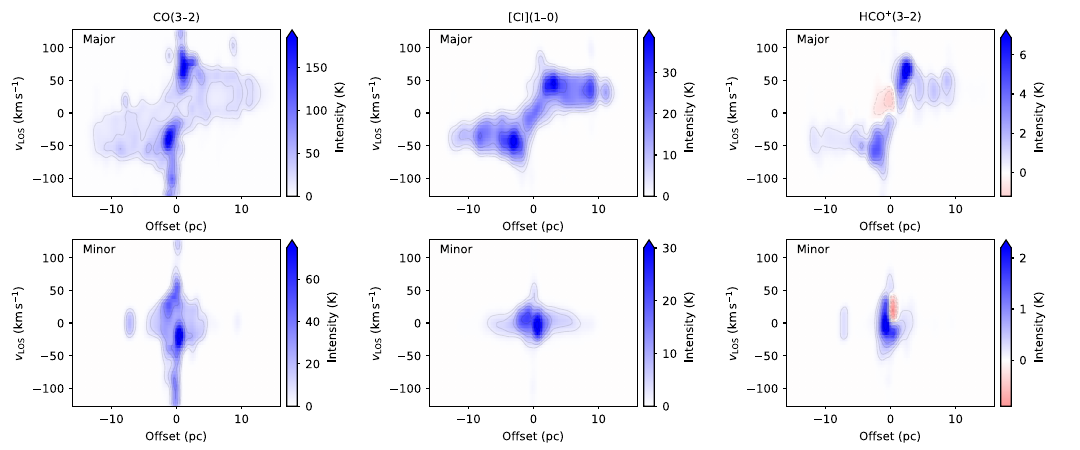}
\caption{
Major and minor axis PVDs of CO(3--2), [\ion{C}{1}](1--0), and HCO$^+$(3--2) observed from $(i,\varphi)=(80^\circ,0^\circ)$.
The contours are drawn at $-$0.3, $-$0.2, $-$0.1, $-$0.05, 0.05, 0.1, 0.2, 0.3, 0.5, 0.7, and 0.9 times the maximum of each panel.
The slit width used to measure the average intensity at each offset and velocity is 1\,pc.
The diagrams are also blurred along the slit with a Gaussian kernel of 1\,pc FWHM.
The CO(3--2) major axis PVD and the [\ion{C}{1}](1--0) minor axis PVD are identical to Figures \ref{fig:CO_PVDs}(b) and \ref{fig:CI_PVDs}(b), respectively.
\label{fig:allPVDs}}
\end{figure*}

From the same spectral cube shown in Figure \ref{fig:images}, we construct PVDs for all CO(3--2), [\ion{C}{1}](1--0), and HCO$^+$(3--2) lines for both the major and minor axes, as shown in Figure \ref{fig:allPVDs}.
The CO(3--2) major and minor PVDs clearly indicate Keplerian rotation for $r<2$\,pc.
Keplerian rotation is less evident in the [\ion{C}{1}](1--0) PVDs.
The minor-axis PVD of [\ion{C}{1}](1--0) shows a polar elongation to $r\sim7$\,pc, in contrast to the other two lines, by tracing the biconical distribution of the C$^0$ gas (upper middle panel of Figure \ref{fig:denities}).
The HCO$^+$(3--2) PVDs exhibit a Keplerian rotation.
Continuum absorption is also evident in these PVDs.
The pattern of the major-axis PVD consists of several blobs in the outer region, reflecting an inhomogeneous HCO$^+$ distribution (see details in Section \ref{sec:profiles}).

\section{Evolution of an HCO$^+$ Line Profile} \label{sec:movie}

\begin{figure*}[tp]
\begin{interactive}{animation}{profile_evolution.mp4}
\plotone{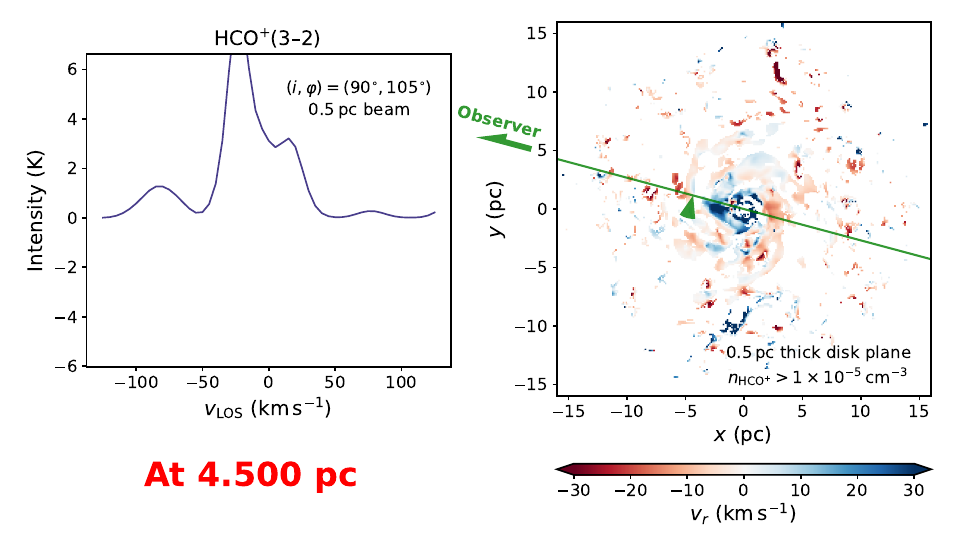}
\end{interactive}
\caption{
HCO$^+$(3--2) line profile obtained while traveling along the LOS through the model.
The left and right panels show a line profile and the radial velocity of HCO$^+$ clumps in the disk plane, corresponding to Figures \ref{fig:origins}(a) and (d), respectively, but here the profile is not the final but a transitional one obtained at an intermediate position of the LOS coordinate $+$4.5\,pc, and the clump velocities are shown for the full range of the model ($r<16$\,pc).
The LOS direction is drawn as the green line in the right panel, and the current position on the LOS is indicated by the green wedge.
An animated version of this figure is available.
The animation shows how the profile evolves and reaches inverse P-Cygni over the entire range of the LOS.
\label{fig:movie}}
\end{figure*}

In Figure \ref{fig:origins}, we selected the velocity channels of the emission and absorption peaks in the final line profile, plotted the intensity growth curves in these channels, and associated them with the gas distribution in the model.
Here we show how the entire profile (all velocity channels) evolves along the LOS as an animation in Figure \ref{fig:movie}.
This animation clearly illustrates the changes described in Section \ref{sec:profiles}.
This animation also shows the origins of other features.
For example, the secondary emission peak ($-$80\,\kmps) occurs at the near-side outflow at $r\leq 1.5$\,pc.

%% Reference
%% BibTeX plus aasjournals.bst to generate the bibliography.
\bibliography{reference}{}
\bibliographystyle{aasjournal}

%% Full list of authors
%% This command is needed when the collaboration and author truncation commands are used.

%% Changes

\end{document}